\documentclass[12pt]{article}
\usepackage{setspace}
\usepackage{epsfig}
\usepackage{amsmath}
\usepackage{amssymb}
\usepackage{graphicx}
\usepackage{lscape}
\usepackage {subfigure}

\def\be{\begin{equation}}
\def\ee{\end{equation}}
\def\ba{\begin{array}}
\def\ea{\end{array}}
\def\beqn{\begin{eqnarray}}
\def\eeqn{\end{eqnarray}}

\def\bt{\begin{tabular}}
\def\et{\end{tabular}}
\def\bc{\begin{center}}
\def\ec{\end{center}}
\begin{document}
\title{Revisiting the texture zero neutrino mass matrices}
\author{Madan Singh, Gulsheen Ahuja$^*$, Manmohan Gupta \\
{\it Department of Physics, Panjab University,
 Chandigarh, India.}\\
\\{\it $^*$gulsheen@pu.ac.in}}

\maketitle

\begin{abstract}
In the light of refined and large reactor mixing
angle $\theta_{13}$, we have revisited the
texture three and two zero neutrino mass matrices
in the flavor basis. For Majorana neutrinos, it
has been explicitly shown that all the texture
three zero mass matrices remain ruled out.
Further, for both normal and inverted mass
ordering, for the texture two zero neutrino mass
matrices one finds interesting constraints on the
Dirac-like CP violating phase $\delta$ and
Majorana phases $\rho$ and $\sigma$.
\end{abstract}

The recent measurement of reactor mixing angle
$\theta_{13}$ and its subsequent refinements
\cite{th13-1}-\cite{th13-4}, along with the
precision measurement of the solar and
atmospheric mixing angles $\theta_{12}$ and
$\theta_{23}$ as well as of the neutrino mass
squared differences have given a new impetus to
the neutrino oscillation phenomenology. The
observation of non zero value of $\theta_{13}$,
on the one hand, restored the parallelism between
quark and lepton mixing, while on the other hand,
it has triggered great deal of interest in the
exploration of CP violation in the leptonic
sector. It has also brought into focus the issue
of constraining the CP violating phases of the
mixing matrix from the textures of the mass
matrices. This has further been strengthened by a
very recent constraint on the sum of absolute
neutrino masses provided by Planck experiment
\cite{planck}. In particular, efforts have been
put in to carry out fine grained analysis
regarding the compatibility of texture specific
mass matrices in the flavor \cite{flavor} as well
as non flavor basis \cite{nonflavor1,
nonflavor2}. Specifically, in the flavor basis
wherein the charged lepton mass matrix is
considered to be diagonal, good deal of attempts
have been made to explore the compatibility of
texture zero mass matrices for Majorana neutrinos
with the neutrino oscillation data.

Present refinements in the data warrant a re-look
at the compatability of the texture zero neutrino
mass matrices. Therefore, the purpose of the
present paper is to re-investigate all the
possible structures of texture three and two zero
neutrino mass matrices in the flavor basis. In
particular, we have carried out an analysis,
similar to the one carried out by Ref.
\cite{xing2011}, with an emphasis on obtaining
useful constraints on the CP violating phases.
Specifically, for the viable possibilities of
texture zero mass matrices, we have examined the
implications of relatively large and refined
$\theta_{13}$ on the parameter space of the
Dirac-like CP violating phase $\delta$ and the
Majorana phases $\rho$, $\sigma$.

The plan of the paper is as follows. To begin
with, some essential details pertaining to the
construction of the Pontecorvo Maki Nakagawa
Sakata (PMNS) matrix \cite{pmns} from the mass
matrices have been presented in Section
(\ref{meth}). Using the inputs given in Section
(\ref{inputs}) and keeping focus on the CP
violating phases $\rho, \sigma$, $\delta$, the
results pertaining to the analyses of texture
three zero Majorana mass matrices have been
presented in Section (\ref{tex3}). For the
texture two zero case, results of the analysis
for Majorana neutrino mass matrices have been
given in Section (\ref{tex2maj}). Finally,
Section (\ref{conc}) summarizes our conclusions.

\section{Texture specific mass matrices and construction of the
PMNS matrix}\label{meth} Before proceeding
further, we briefly underline the methodology
relating the elements of the mass matrices to
those of the mixing matrix. In the flavor basis,
wherein the charged lepton mass matrix $M_{l}$ is
diagonal, the Majorana neutrino mass matrix
$M_{\nu}$ can be expressed in terms of three
neutrino masses $m_{1}$,$m_{2}$, $m_{3}$ and the
flavor mixing matrix V as
\begin{equation}
 M_{\nu}=V\left(
\begin{array}{ccc}
    m_{1}& 0& 0 \\
  0 & m_{2} & 0\\
  0& 0& m_{3} \\
  \end{array}
  \right)V^{T}.
\end{equation}
The mixing matrix V can be written as $V=UP$,
where $U$ denotes the Pontecorvo-Maki-Nakagawa
Sakata (PMNS) \cite{pmns} neutrino mixing matrix
consisting of three flavor mixing angles and one
Dirac-like CP violating phase, whereas, the
matrix $P$ is a diagonal phase matrix, i.e.,
$P$=diag($e^{i\rho},e^{i\sigma},1$) with $\rho$
and $\sigma$ being the two Majorana CP violating
phases. The neutrino mass matrix $M_{\nu}$ can
then be rewritten as
\begin{equation}
M_{\nu}=\left(
\begin{array}{ccc}
   M_{ee}& M_{e \mu}& M_{e \tau} \\
  M_{e \mu} & M_{\mu \mu} & M_{\mu \tau}\\
  M_{e \tau}& M_{\mu \tau}& M_{\tau \tau} \\
  \end{array}
  \right)=U\left(
\begin{array}{ccc}
    \lambda_{1}& 0& 0 \\
  0 & \lambda_{2} & 0\\
  0& 0& \lambda_{3} \\
\end{array}
\right) U^{T},
\end{equation}\\
where $\lambda_{1} = m_{1} e^{2i\rho},\lambda_{2}
= m_{2} e^{2i\sigma} ,\lambda_{3} = m_{3}.$

For the purpose of calculations, we have adopted
the parameterization of the mixing matrix $U$
considered by Ref. \cite{xing2011}, e.g.,
\begin{equation}
U=\left(
\begin{array}{ccc}
 c_{12}c_{13}& s_{12}c_{13}& s_{13} \\
-c_{12}s_{23}s_{13}-s_{12}c_{23}e^{-i\delta} &
-s_{12}s_{23}s_{13}+c_{12}c_{23}e^{-i\delta} &
s_{23}c_{13}\\
 -c_{12}c_{23}s_{13}+s_{12}s_{23}e^{-i\delta}
 & -s_{12}c_{23}s_{13}-c_{12}s_{23}e^{-i\delta}& c_{23}c_{13} \\
\end{array}
 \right),
\end{equation}\\
where $c_{ij} = cos \theta_{ij}, s_{ij}= sin
\theta_{ij}$ for i,j=1,2,3 and $\delta$ is the CP
violating phase.

\section{Inputs used in the present analysis}\label{inputs}
Before discussing the results of the analysis, we
summarize the experimental information about
various neutrino oscillation parameters. For both
normal mass ordering (NO) and inverted mass
ordering (IO), the best fit values and the latest
experimental constraints on neutrino parameters
at 1$\sigma$, 2$\sigma$ and 3$\sigma$ confidence
level (CL), following Ref. \cite{forero}, are
given in Table (\ref{data}).
\begin{table}[hb]
\begin{small}
\begin{center}
\begin{tabular}{|c|c|c|c|c|}
\hline
Parameter& Best Fit & 1$\sigma$ & 2$\sigma$ &
3$\sigma$ \\ \hline $\delta m^{2}$
$[10^{-5}eV^{2}]$ & $7.60$& $7.42$ - $7.79$ &
$7.26$ - $7.99$ & $7.11$ - $8.18$ \\ \hline
$|\Delta m^{2}_{31}|$ $[10^{-3}eV^{2}]$ (NO) &
$2.48$ & $2.41$ - $2.53$ & $2.35$ - $2.59$ &
$2.30$ - $2.65$\\ \hline $|\Delta m^{2}_{31}|$
$[10^{-3}eV^{2}]$ (IO) & $2.38$ & $2.32$ - $2.43$
& $2.26$ - $2.48$ & $2.20$ - $2.54$ \\ \hline
$\theta_{12}$ & $34.6^{\circ}$ & $33.6^{\circ}$ -
$35.6^{\circ}$ & $32.7^{\circ}$ - $36.7^{\circ}$
& $31.8^{\circ}$ - $37.8^{\circ}$\\ \hline $
\theta_{23}$ (NO) & $48.9^{\circ}$
&$41.7^{\circ}$ - $50.7^{\circ}$  &
$40.0^{\circ}$ - $52.1^{\circ}$ & $38.8^{\circ}$
- $53.3^{\circ}$ \\ \hline $\theta_{23}$ (IO)&
$49.2^{\circ}$ & $46.9^{\circ}$ - $50.7^{\circ}$
& $41.3^{\circ}$ - $52.0^{\circ} $&
$39.4^{\circ}$ - $53.1^{\circ}$ \\ \hline
$\theta_{13}$ (NO) & $8.6^{\circ}$ &
$8.4^{\circ}$ - $8.9^{\circ}$ & $8.2^{\circ}$ -
$9.1^{\circ}$& $7.9^{\circ}$ - $9.3^{\circ}$ \\
\hline $\theta_{13}$ (IO) & $8.7^{\circ}$ &
$8.5^{\circ}$ - $8.9^{\circ}$ & $8.2^{\circ}$ -
$9.1^{\circ}$ & $8.0^{\circ}$ - $9.4^{\circ}$ \\
\hline $\delta$ (NO) & $254^{\circ}$ &
$182^{\circ}$ - $353^{\circ}$& $0^{\circ}$ -
$360^{\circ}$ & $0^{\circ}$ - $360^{\circ}$ \\
\hline $\delta$ (IO) &$266^{\circ}$&
$210^{\circ}$ - $322^{\circ}$ & $0^{\circ}$ -
$16^{\circ}$ $\oplus$ $ 155^{\circ}$ -
$360^{\circ}$  & $0^{\circ}$ - $360^{\circ}$ \\
\hline
\end{tabular}
\caption{\label{data}1$\sigma$, 2$\sigma$ and
3$\sigma$ CL ranges of neutrino oscillation
parameters. NO (IO) refers to normal (inverted)
neutrino mass ordering.}
\end{center}
\end{small}
\end{table}

\section{Texture three zero Majorana neutrino mass
matrices}\label{tex3} In the flavor basis, there
are 20 possible texture three zero patterns, the
texture structures of these have been elaborated
in Ref \cite{xing2004} and it has been shown that
all these are found to be incompatible with the
neutrino oscillation data. Our present analysis,
carried out be the current refined data also
reinforces this earlier conclusion. To elaborate
this, we explicitly examine one of the
possibilities and bring forward its
incompatibility with the latest neutrino mixing
data. In particular, we consider the case wherein
$M_{ee}$=0, $M_{e \tau}$=0, $M_{\mu\mu}$=0, i.e.,
\begin{equation}
M_{\nu}=\left(
\begin{array}{ccc}
    0& \times & 0\\
  \times & 0 & \times\\
  0& \times & \times \\
\end{array}
\right).
\end{equation}

The neutrino mass matrix $M_{\nu}$ can be
diagonalized  by using flavor mixing matrix U.
For the present case, the constraints $M_{e e}$=0
and $M_{e \tau}$=0 yields the following
expression for ratios of the complex neutrino
mass eigenvalues
\begin{equation}
\frac{\lambda_{1}}{\lambda_{3}}= -\dfrac{s_{13}}{c_{13}^{2}}
  \bigg(\frac{s_{12}c_{23}}{c_{12}s_{23}}e^{i\delta}+s_{13}\bigg);
  \qquad \frac{\lambda_{2}}{\lambda_{3}}= +\dfrac{s_{13}}{c_{13}^{2}}
  \bigg(\frac{c_{12}c_{23}}{s_{12}s_{23}}e^{i\delta}-s_{13}\bigg).
\end{equation}
For $M_{\mu\mu}$=0, we obtain
\begin{equation}
\lambda_{3}\bigg(U_{\mu 1}U_{\mu
1}.\frac{\lambda_{1}} {\lambda_{3}}+U_{\mu
2}U_{\mu
2}.\frac{\lambda_{2}}{\lambda_{3}}+U_{\mu
3}U_{\mu 3}\bigg)=0.
 \end{equation}
From the real and imaginary parts of the above
equations, we obtain the following constraints on
$\theta_{12}$. From the real part, we get
\begin{equation}
\theta_{12}=\frac{1}{2}
acot\bigg(\frac{s_{23}c_{23}^{2}s_{13}^{2}
(2+cot2\delta)-s_{23}^{3}cos2\theta_{13}}
{2c_{23}^{2}.s_{13}cos\delta}\bigg),
\end{equation}
whereas, from the imaginary part, we obtain
\begin{equation}
\theta_{12}=\frac{1}{2} acot\bigg(\frac{s_{13}
cos(2\delta)s_{23}}{2sin\delta}\bigg).
\end{equation}

\begin{figure}[ht]
\begin{center}
\subfigure[]{\includegraphics[width=0.45\columnwidth]{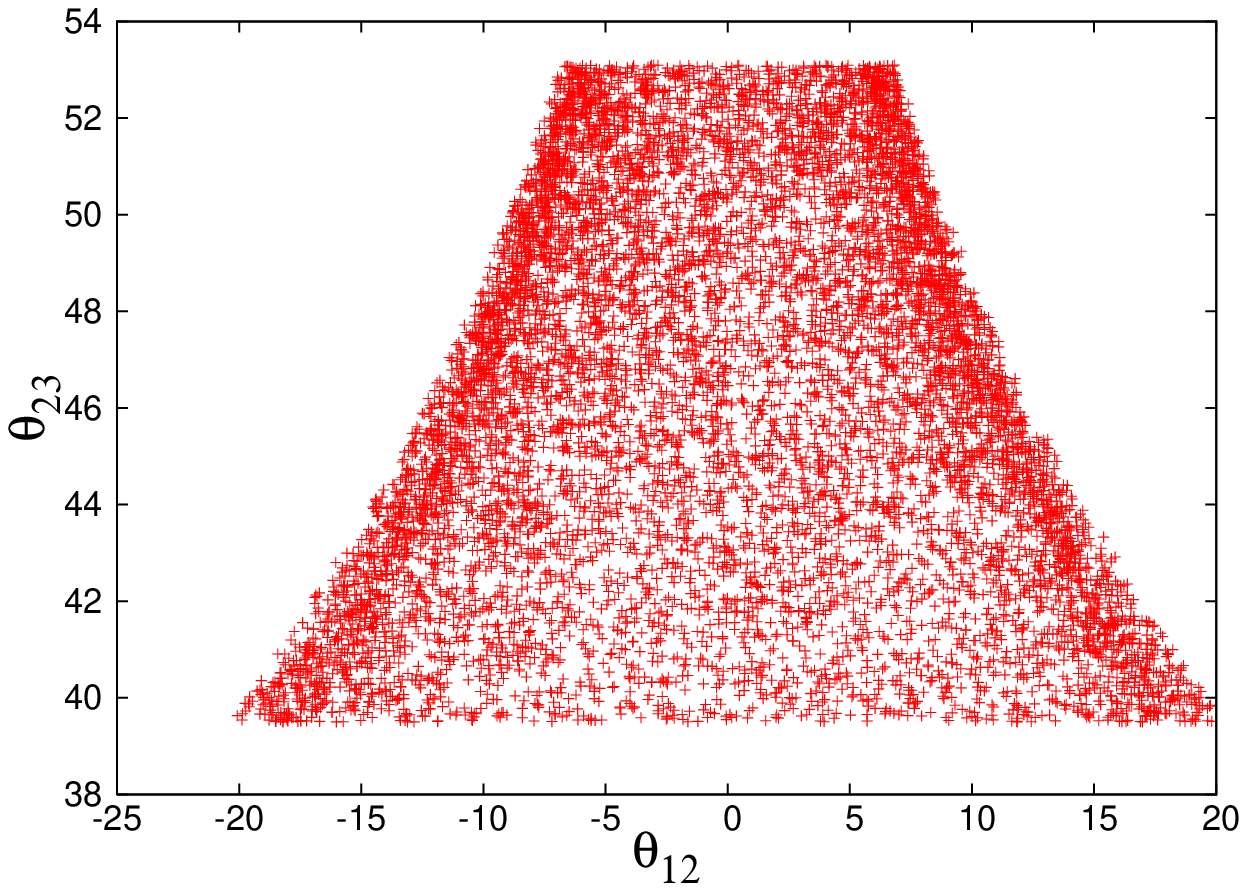}}
\ \ \
\subfigure[]{\includegraphics[width=0.45\columnwidth]{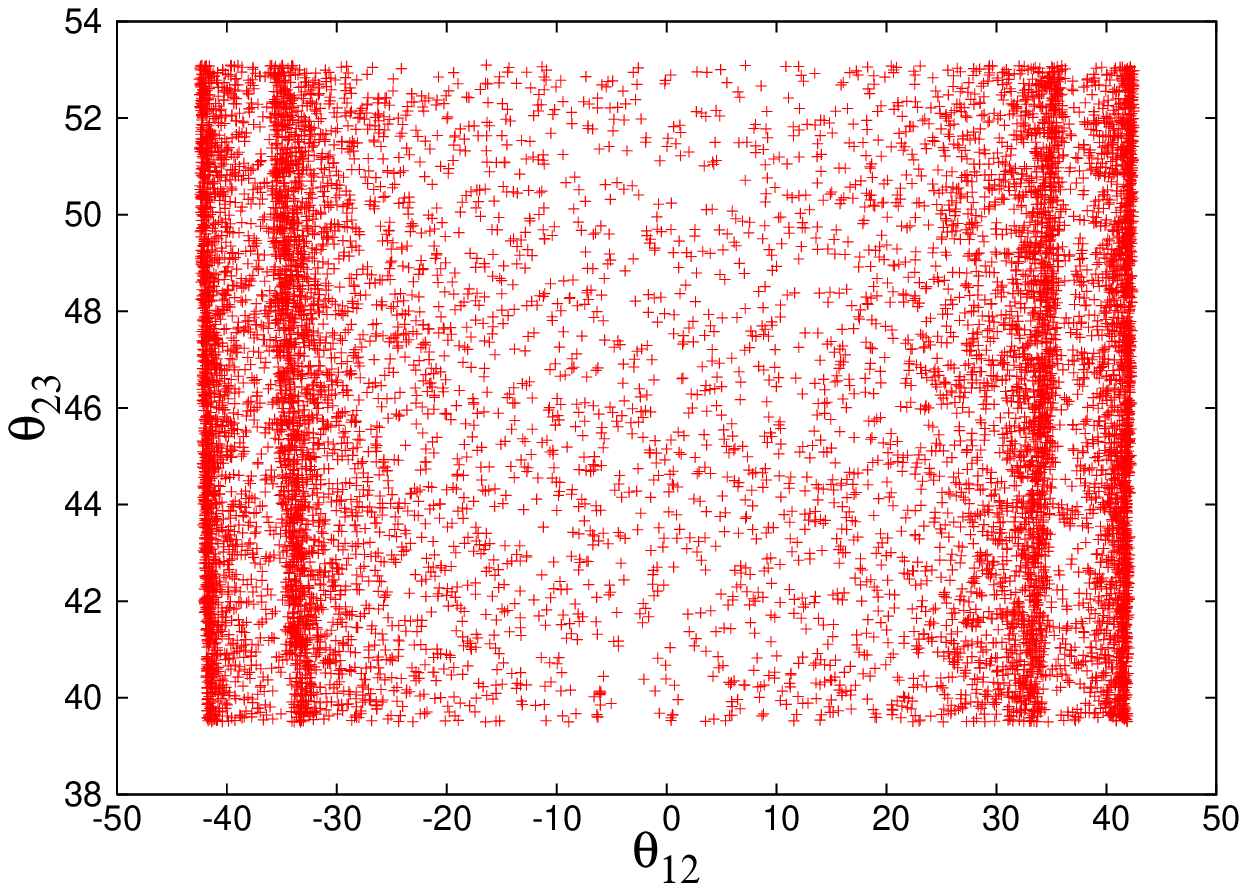}}\\
\caption{\label{fig1} Correlation plots for
(a)real and (b) imaginary part of $M_{\mu \mu}$=0
for three- zero texture with $M_{ee}$=0, $M_{e
\tau}$=0, $M_{\mu\mu}$=0, where angles are
measured in degrees.}
\end{center}
\end{figure}

To check the compatibility of the above two
solutions, in Fig.\ref{fig1}(a) and (b), we have
presented the correlation plots between
$\theta_{12}$ and $\theta_{23}$ for the
constraints obtained from the real and imaginary
parts respectively. These graphs have been
obtained by giving full allowed variation to
phase $\delta$ and 3$\sigma$ variation to
$\theta_{13}$.
From Fig.\ref{fig1}(a), we find that the range
obtained for $\theta_{12}$ here has no overlap
with its experimental range, i.e., $31.8^{0}$
-$37.8^{0}$, whereas, the plot given in
Fig.\ref{fig1}(b) shows that $\theta_{12}$ found
here includes its experimental range. For the
compatibility of the texture three zero
possibility considered here, both the graphs
should include the experimental range of
$\theta_{12}$, therefore, ruling out the present
case. A similar analysis corresponding to the
other possibilities of texture three zero
Majorana neutrino mass matrices also yield the
same result, therefore ruling these out.

\section{Texture two zero Majorana neutrino mass
matrices}\label{tex2maj} Coming to the case of
texture two zero Majorana neutrino mass matrices,
it is well known that there are 15 possible
structures of neutrino mass matrix. Adopting the
classification scheme of Ref. \cite{xing2011},
these can be divided into six classes A, B, C, D,
E and F. For the purpose of examining whether a
possible pattern of texture two zero mass
matrices is viable or not, one needs to examine
the compatability of the corresponding PMNS
matrix constructed using these. For executing the
present analysis, to begin with, we vary the
input neutrino oscillation parameters, i.e., the
three mixing angles $\theta_{12}$, $\theta_{23}$
and $\theta_{13}$ as well as the mass squared
differences $\delta m^{2}$ and $\Delta m^{2}$
within their $3\sigma$ CL ranges, summarized in
Table 3.1. In the absence of any experimental
constraint on Dirac-like CP violating phase
$\delta$, we give full variation of $0^\circ$ to
$360^{\circ}$ to it. In order to obtain
constraints on the Majorana phases $\rho$ and
$\sigma$, these can be expressed in terms of the
mixing matrix elements, e.g.,
\begin{equation}
\rho=\dfrac{1}{2}arg
\left(\frac{U_{a3}U_{b3}U_{l2}U_{m2}-U_{a2}
U_{b2}U_{l3}U_{m3}}{U_{a2}U_{b2}U_{l1}U_{m1}-U_{a1}U_{b1}U_{l2}U_{m2}}\right),
\end{equation}
and
\begin{equation}
\sigma=\dfrac{1}{2}arg
\left(\frac{U_{a1}U_{b1}U_{l3}U_{m3}-U_{a3}
U_{b3}U_{l1}U_{m1}}{U_{a2}U_{b2}U_{l1}U_{m1}-U_{a1}U_{b1}U_{l2}U_{m2}}\right).
\end{equation}

Coming to the results of the analysis, it may be
mentioned that even with the present refinements
of the mixing angle $\theta_{13}$, out of the 15
possible texture two zero Majorana neutrino mass
matrices, the texture patterns $A_{1,2}$,
$B_{1,2,3,4}$ and C remain viable for the normal
mass ordering (NO). The possibilities $D_{1, 2}$
have already been ruled out by Ref.
\cite{xing2011} due to mixing angle $\theta_{12}$
being less than $38^{\circ}$, conclusions of our
analysis in this regard remain the same.
Similarly, patterns $ E_{1,2,3}$ and $F_{1,2,3}$
remain phenomenologically disfavored. In the
following, for NO, we present results for the
patterns of categories A, B and C. In particular,
we present the implications of the relatively
large and refined $\theta_{13}$ on the parameter
space of three CP violating phases $\delta$,
$\rho$ and $\sigma$.

\subsection{Class A}
\begin{figure}[ht]
\begin{center}
\includegraphics[width=0.40\columnwidth]{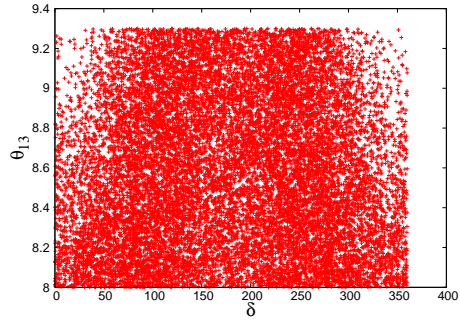}
\caption{Pattern $A_{1}$: Mixing angle
$\theta_{13}$ versus phase $\delta$. All the
parameters are in degrees.} \label{figA1}
\end{center}
\end{figure}

This class consists of 2 possible texture two
zero matrices $A_{1}$ and $A_{2}$. Interestingly,
there exists a 2-3 permutation symmetry between
these, therefore, the phenomenological
implications of these are similar and thus we
have discussed pattern $A_{1}$ only in detail. As
a first step, to examine the implications of non
zero and large $\theta_{13}$ on phase $\delta$,
in Fig.\ref{figA1}, we have given $\theta_{13}$
versus $\delta$ plot. From the graph, one finds
that corresponding to the 3$\sigma$ CL
experimental range of angle $\theta_{13}$ i.e.
$7.9^\circ$ to $9.3^{\circ}$, one does not obtain
any constraint on phase $\delta$, this being in
agreement with the conclusions of
Ref.\cite{xing2011}.

\begin{figure}[ht]
\begin{center}
\subfigure[]{\includegraphics[width=0.40\columnwidth]{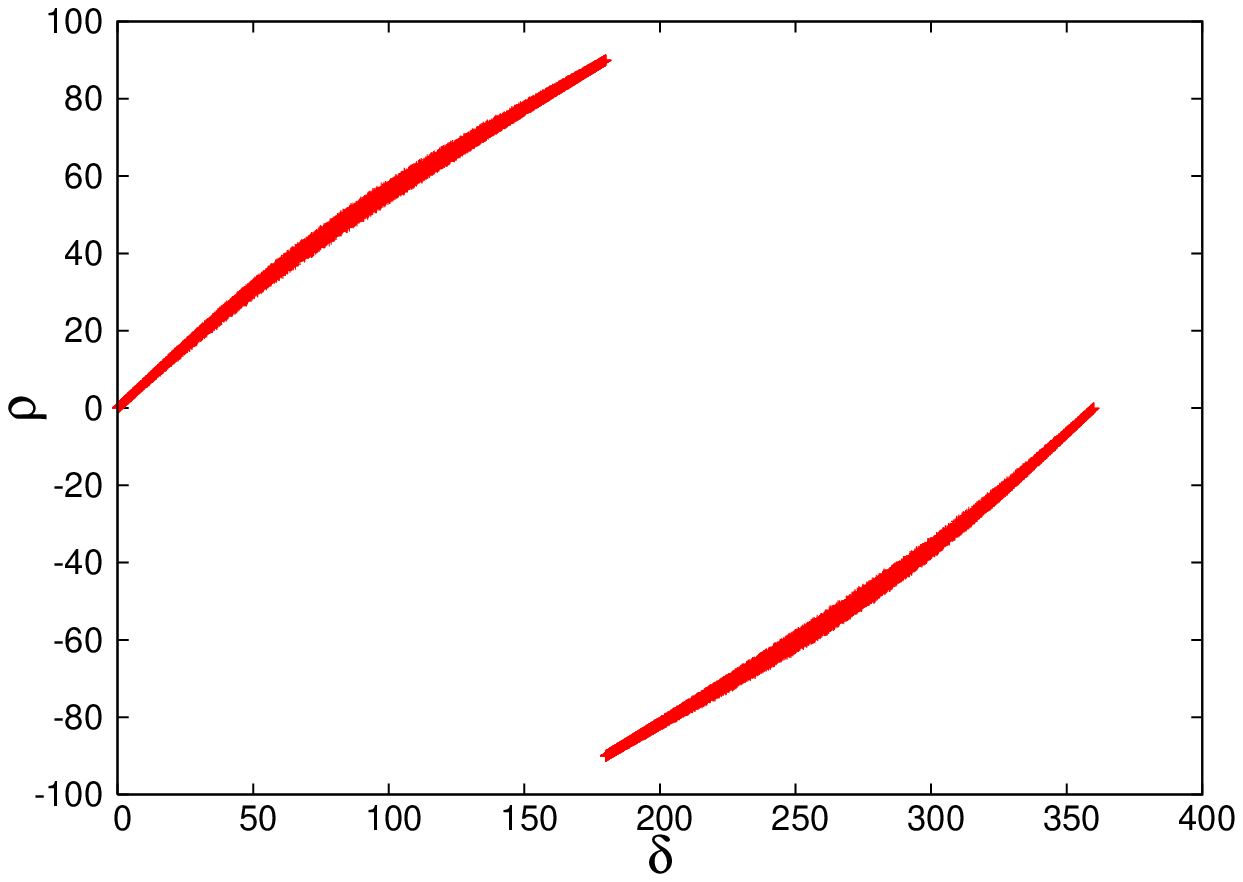}}
\ \ \
\subfigure[]{\includegraphics[width=0.40\columnwidth]{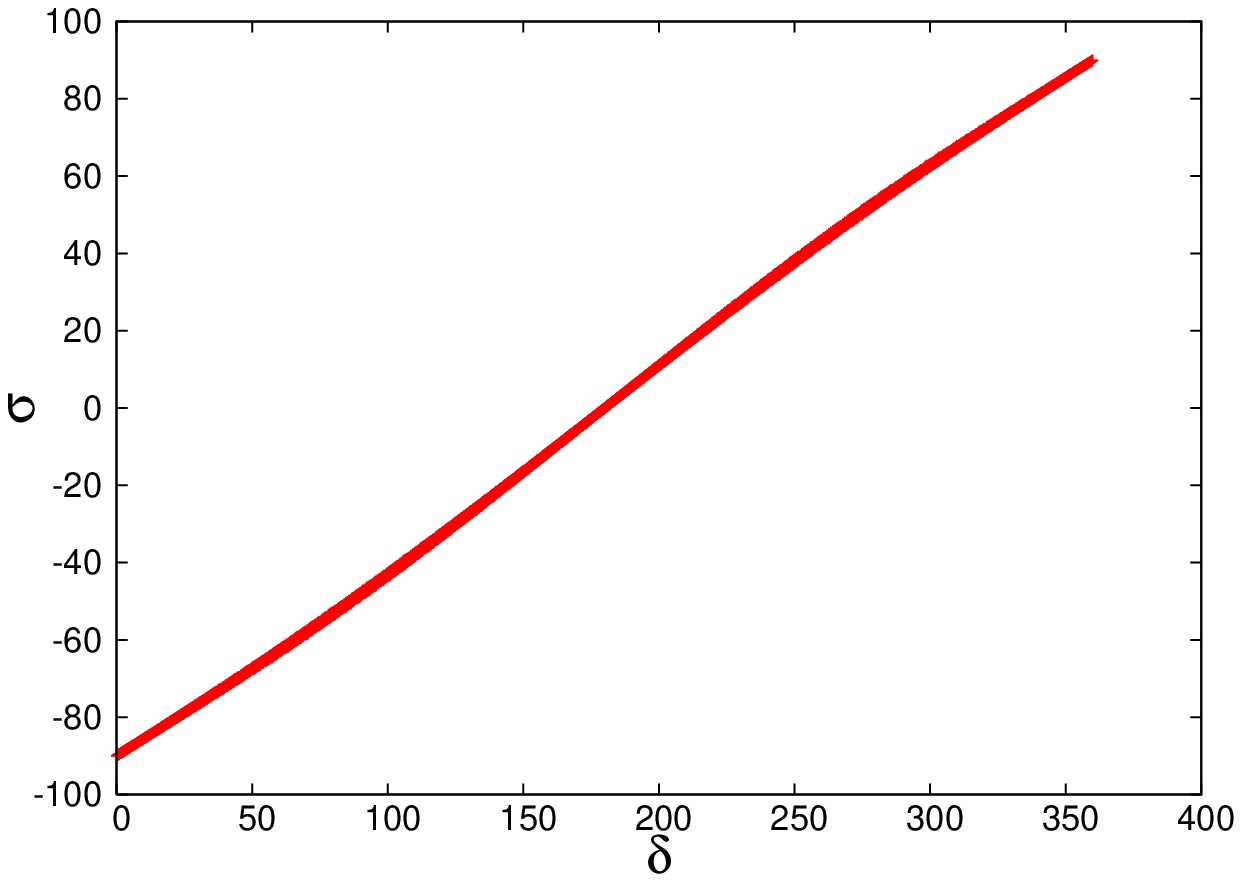}}\\
\caption{\label{figAphases} Pattern $A_{1}$:
Correlation plots of phases $\rho$ and $\sigma$
versus $\delta$. All the parameters are in
degrees.}
\end{center}
\end{figure}

As a next step, in Fig.\ref{figAphases}(a) and
(b) we have given the correlation plots of
Dirac-like CP violating phase $\delta$ and
Majorana phases $\rho$ and $\sigma$. Again, we
find that one is not able to obtain any useful
constraints on these phases, in particular, both
$\rho$ and $\sigma$ take values from
$-90^{0}-90^{0}$.

\subsection{Class B}
\begin{figure}[ht]
\begin{center}
\includegraphics[width=0.40\columnwidth]{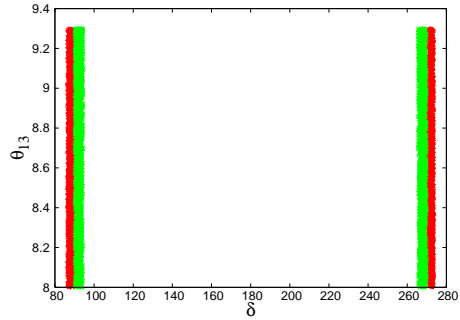}
 \caption{Patterns $B_{1}$ (in
green) and $B_{3}$ (in red): Mixing angle
$\theta_{13}$ versus phase $\delta$. All the
parameters are in degrees.} \label{fig2}
\end{center}
\end{figure}

The class B consists of 4 possible texture two
zero matrices $B_{1, 2, 3, 4}$. Similar to the
matrices of class A, there exists a 2-3
permutation symmetry between the matrices $B_{1}$
and $B_{2}$ as well as between $B_{3}$ and
$B_{4}$, therefore, the phenomenological
implications of these are similar and thus we
have discussed the patterns $B_{1}$ and $B_{3}$
only in detail. In order to obtain constraints on
Dirac-like CP violating phase $\delta$ due to the
refined measurement of mixing angle
$\theta_{13}$, in Fig. \ref{fig2}, we have given
$\theta_{13}$ versus $\delta$ plot for patterns
$B_{1}$ and $B_{3}$. From the graph one finds
that unlike the case of matrices belonging to
class A, the 3$\sigma$ CL experimental range of
angle $\theta_{13}$ considerably shrinks the
parameter space of $\delta$ near $90^{0}$ and
$270^{0}$. Therefore, further refinements in the
measurement of phase $\delta$ would have
implications for these matrices considered here.

\begin{figure}[ht]
\begin{center}
\subfigure[]{\includegraphics[width=0.40\columnwidth]{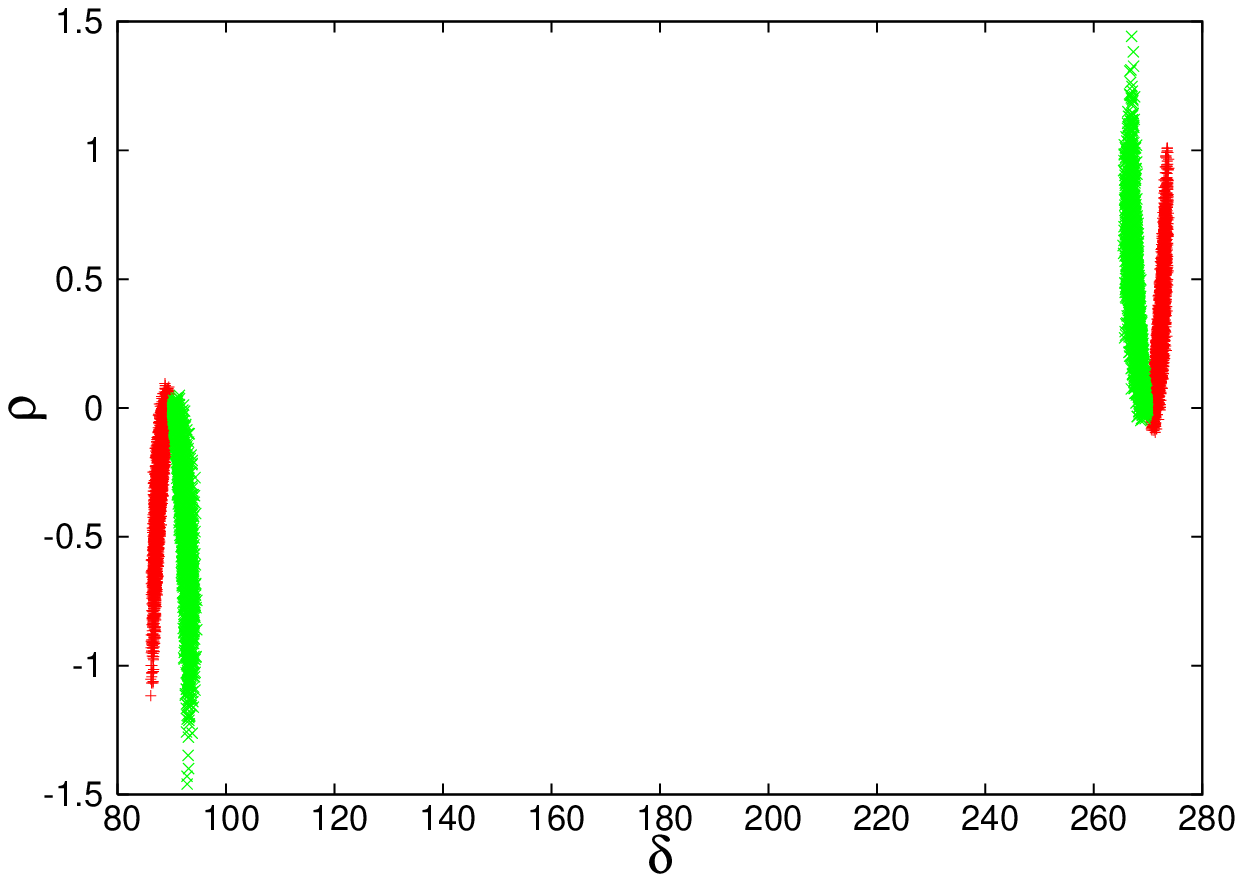}}
\ \ \
\subfigure[]{\includegraphics[width=0.40\columnwidth]{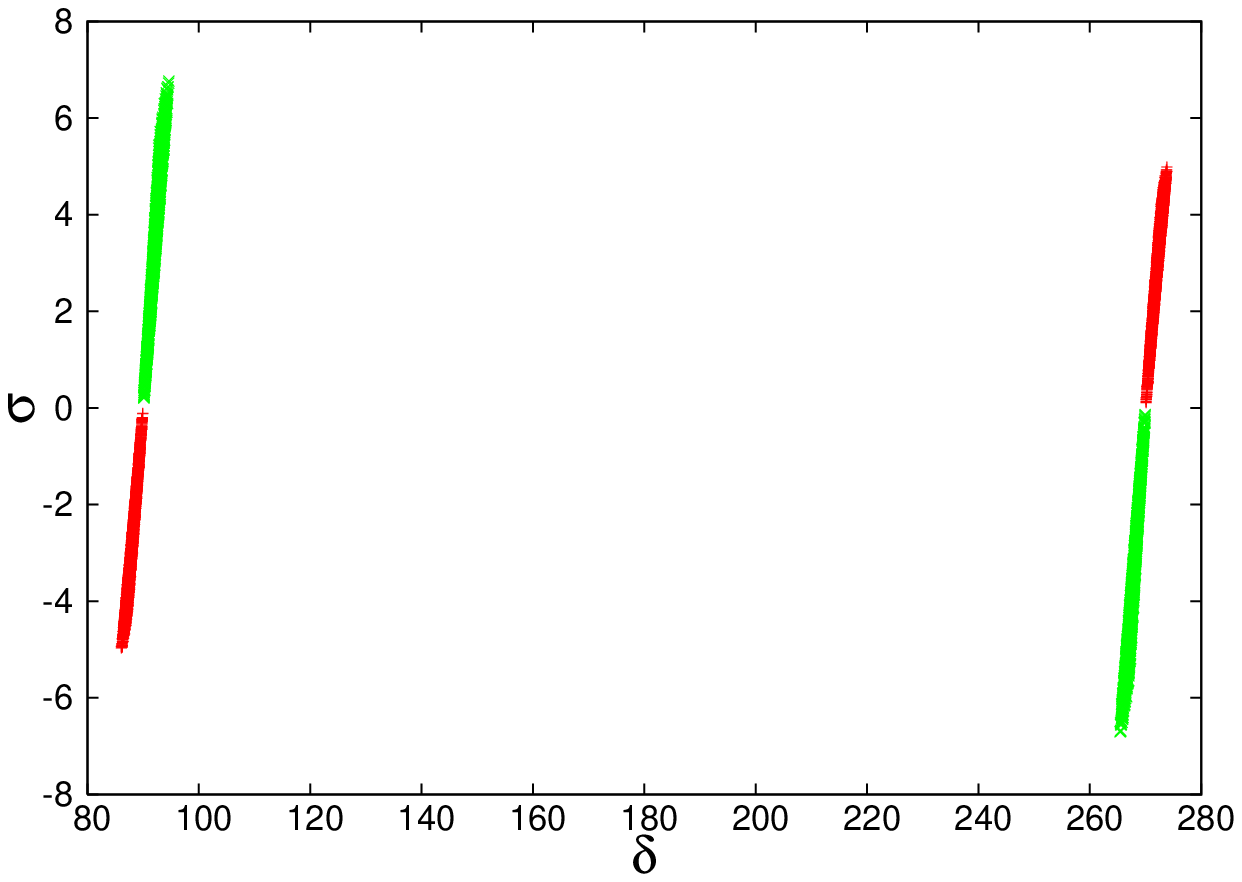}}\\
\caption{\label{fig3} Patterns $B_{1}$ (in green)
and $B_{3}$ (in red): Correlation plots of phases
$\rho$ and $\sigma$ versus $\delta$. All the
parameters are in degrees.}
\end{center}
\end{figure}

For the matrices $B_{1}$ and $B_{3}$, the
correlation plots of Dirac-like CP violating
phase $\delta$ and Majorana phases phases $\rho$
and $\sigma$ have been shown in Fig.
\ref{fig3}(a) and (b). The plots indicate that
the Majorana phases $\rho$ and $\sigma$ now take
quite small values, in particular, phase $\rho$
lies between $-1.5^{0}-1.5^{0}$ , whereas,
$\sigma$ take values from $-7^{0}-7^{0}$.

\subsection{Class C}
\begin{figure}[ht]
\begin{center}
\includegraphics[width=0.40\columnwidth]{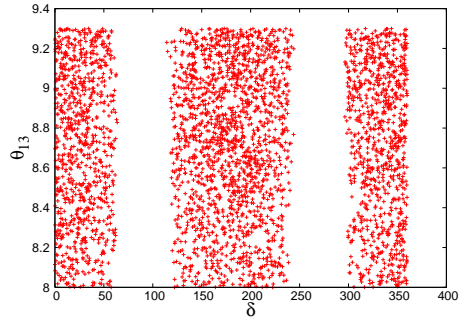}
\caption{Pattern $C$: Mixing angle $\theta_{13}$
versus phase $\delta$. All the parameters are in
degrees.} \label{figC}
\end{center}
\end{figure}

Similar to the earlier classes, for the matrix
belonging to class C, as a first step, we have
examined the implications of non zero and large
$\theta_{13}$ on phase $\delta$, in Fig.
\ref{figC} we have given $\theta_{13}$ versus
$\delta$ plot. From the graph, one finds that
corresponding to the 3$\sigma$ CL experimental
range of angle $\theta_{13}$, the phase $\delta$
takes values from $0^{0}-60^{0}$,
$120^{0}-230^{0}$ and $300^{0}-350^{0}$ .

\begin{figure}[ht]
\begin{center}
\subfigure[]{\includegraphics[width=0.40\columnwidth]{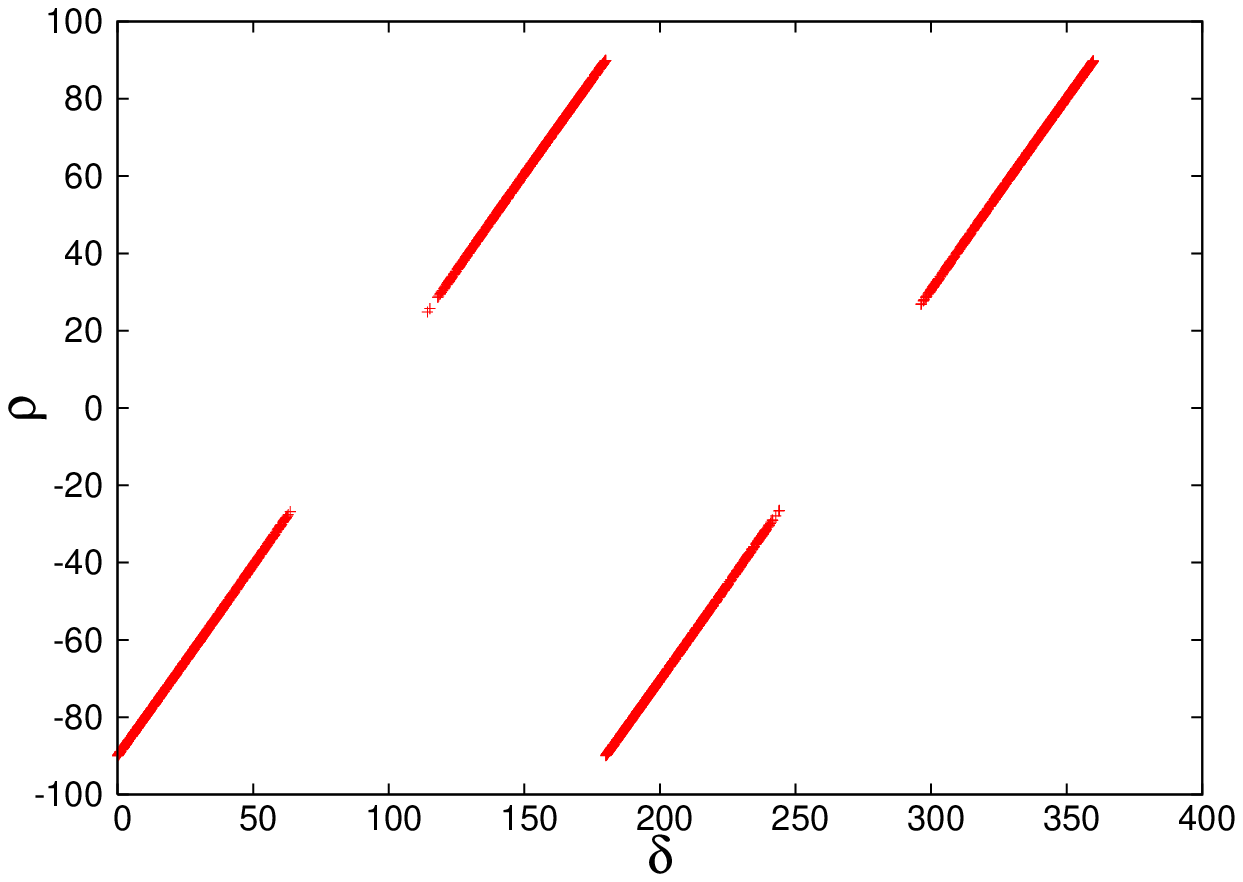}}
\ \ \
\subfigure[]{\includegraphics[width=0.40\columnwidth]{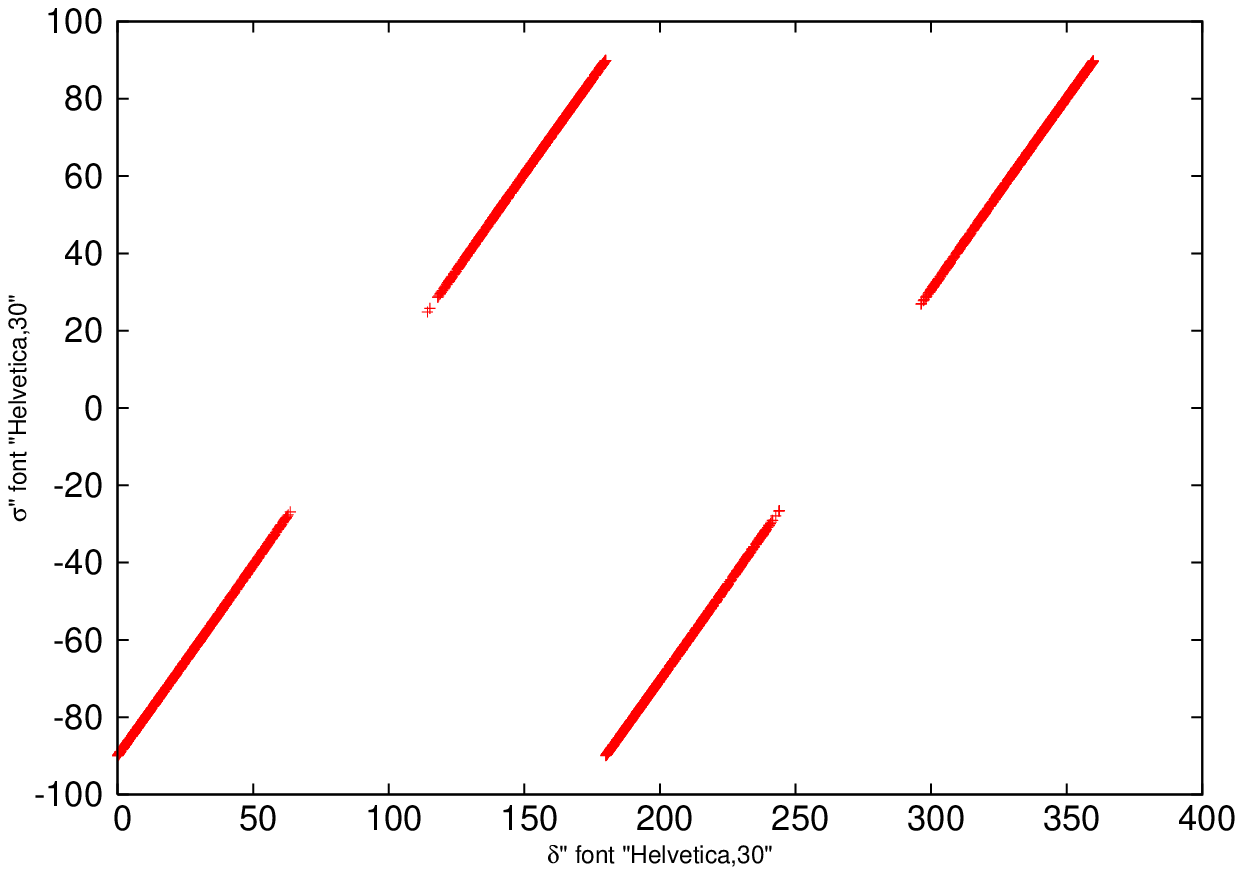}}
\caption{\label{figCphases} Pattern $C$:
Correlation plots of phases $\rho$ and $\sigma$
versus $\delta$. All the parameters are in
degrees.}
\end{center}
\end{figure}

In Fig. \ref{figCphases}(a) and (b) we have given
the correlation plots of Dirac-like CP violating
phase $\delta$ and Majorana phases phases $\rho$
and $\sigma$. For both these phases, we do not
obtain any useful constraint, i.e., these takes
values from $-90^{0}-90^{0}$.

\subsection{Comparing the results for normal and inverted
mass ordering} After having discussed the
implications of recent refinements of neutrino
oscillation parameters on the Dirac-like CP
violating phase $\delta$ as well as on Majorana
phases $\rho$ and $\sigma$, we now briefly
discuss and compare the results of the analysis
for the normal (NO) and inverted (IO) mass
orderings.

\begin{landscape}
\begin{table}[ht]
\begin{small}
\begin{center}
\begin{footnotesize}

\begin{tabular}{|c|c|c|c|c|}
  \hline
Class&\multicolumn{2}{c|}{Normal mass ordering
(NO)} &\multicolumn{2}{c|}{Inverted mass ordering
(IO)} \\ \hline $A_{1}$  & $\delta=
0^{0}-360^{0}$ & $m_{1}=0.0023-0.0113$& $\times$
&$\times$  \\ & $\rho= -90^{0}-90^{0}$ &
$m_{2}=0.0087-0.0144$ & $\times $ & $\times$ \\
&$\sigma= -90^{0}-90^{0}$ & $m_{3}=0.0447-0.0565$
& $\times$ & $\times$ \\ \hline $A_{2}$  &
$\delta= 0^{0}-360^{0}$& $m_{1}=0.0021-0.0113$ &
$\times$ &$\times$  \\ & $\rho= -90^{0}-90^{0}$
&$m_{2}=0.0086-0.0144$  & $\times$ & $\times$ \\
&$\sigma= -90^{0}-90^{0}$ &$m_{3}=0.0447-0.0565$
& $\times$ & $\times$ \\ \hline $B_{1}$  &
$\delta= 90.05^{0}-94.67^{0}$ $\oplus$
$265.3^{0}-269.9^{0}$&  $m_{1}=0.0327-0.326$ &
$\delta= 89.5^{0}-91.03^{0}$ $\oplus$
$268.8^{0}-270.4^{0}$ &$m_{1}=0.0540-0.278$ \\ &
$\rho= -1.8^{0}-0.05^{0}$ $\oplus$
$-0.008^{0}-1.5^{0}$ & $m_{2}=0.0349-0.327$&
$\rho= 0.09^{0}-3.0^{0}$ $\oplus$ $-3.0^{0}-
-1.0^{0}$ & $m_{2}=0.0536-0.0811$ \\ & $\sigma=
0.07^{0}-7.2^{0}$ $\oplus$ $-7.4^{0}- -0.1^{0}$ &
$m_{3}=0.0564-0.331$ & $\sigma= -1.3^{0}-0.2^{0}$
$\oplus$ $-0.3^{0}- 1.3^{0}$ &
$m_{3}=0.0290-0.274$ \\ \hline

$B_{2}$  & $\delta= 83.18^{0}-89.7^{0}$ $\oplus$
$270^{0}-277^{0}$& $m_{1}=0.0276-0.15$& $\delta=
89^{0}-90.5^{0}$ $\oplus$ $269.5^{0}-271^{0}$ &
$m_{1}=0.0594-0.31$  \\ & $\rho=
-0.03^{0}-2.8^{0}$ $\oplus$ $-2.8^{0}-0.027^{0}$
& $m_{2}=0.0285-0.16$ & $\rho= -2.6^{0}-
-0.07^{0}$ $\oplus$ $ 0.1^{0}- 2.3^{0}$
&$m_{2}=0.0599-0.31$ \\ & $\sigma=
-10.5^{0}-0.4{0}$ $\oplus$ $ 0.2^{0}- 10.6^{0}$ &
$m_{3}=0.0522-0.13$  & $\sigma= -0.2^{0}-1.3^{0}$
$\oplus$ $-1.2^{0}- 0.2^{0}$ & $m_{3}=0.0339-0.3$
\\ \hline

$B_{3}$  & $\delta= 86.2^{0}-89.9^{0}$ $\oplus$
$270^{0}-273.9^{0}$&  $m_{1}=0.0350-0.326$ &
$\delta= 89.8^{0}-91.82^{0}$ $\oplus$
$268.2^{0}-270.2^{0}$ &$m_{1}=0.054-0.17$ \\ &
$\rho= -2.8^{0}-0.2^{0}$ $\oplus$
$-0.17^{0}-1.1^{0}$ & $m_{2}=0.0381-0.32$& $\rho=
-2.8^{0}- -0.2^{0}$ $\oplus$ $ 0.06^{0}- 2.8^{0}$
&$m_{2}=0.055-0.17$  \\ & $\sigma=
-5.2^{0}--0.06^{0}$ $\oplus$ $ 0.3^{0}- 5.1^{0}$
&$m_{3}=0.0588-0.32$  & $\sigma=
0.12^{0}-3.4^{0}$ $\oplus$ $-3.4^{0}- -0.05^{0}$
&$m_{3}=0.0313-0.16$  \\ \hline

$B_{4}$  & $\delta= 90.1^{0}-95.9^{0}$ $\oplus$
$263.9^{0}-269.9^{0}$&$m_{1}=0.0298-0.22$ &
$\delta= 88.8^{0}-90^{0}$ $\oplus$
$269.9^{0}-271^{0}$ &$m_{1}=0.0563-0.247$   \\ &
$\rho= -0.16^{0}-2.1^{0}$ $\oplus$
$-2.2^{0}-0.1^{0}$ & $m_{2}=0.0303-0.22$ & $\rho=
0.04^{0}-2.3^{0}$ $\oplus$ $-2.0^{0}- -1.8^{0}$ &
$m_{2}=0.0565-0.32$\\ & $\sigma=
0.18^{0}-7.2^{0}$ $\oplus$ $-7.4^{0}- -0.1^{0}$
&$m_{3}=0.0545-0.22$  & $\sigma=
-2.3^{0}--0.09^{0}$ $\oplus$ $ 0.04^{0}- 2.1^{0}$
& $m_{3}=0.0366-0.32$\\ \hline

$C$  & $\delta= 0^{0}-62.4^{0}$ $\oplus$
$114^{0}-244^{0}$ & $m_{1}=0.135-0.32$& $\delta=
40.5^{0}-86.8^{0}$ $\oplus$ $93.2^{0}-266.5^{0}$&
$m_{1}=0.0482-0.32$ \\ & $\oplus$&
$m_{2}=0.130-0.32$&$\oplus$&
$m_{2}=0.0489-0.32$\\ &$294^{0}-360^{0}$&
$m_{3}=0.137-0.32$ &$273.2^{0}-325^{0}$&
$m_{3}=0.0197-0.32$ \\

& $\rho= -90^{0}- -27.3^{0}$ $\oplus$
$27^{0}-90^{0}$  &   & $\rho= -18.4^{0}-18.2^{0}$
&   \\ & $\sigma= -90^{0}- -27.3^{0}$ $\oplus$
$27^{0}-90^{0}$    &  & $\sigma= -90^{0}-90^{0}$
&
\\ \hline

\end{tabular}

\caption{\label{tab2}The allowed ranges of
Dirac-like CP violating phase $\delta$, the
Majorana phases $\rho, \sigma$ and three neutrino
masses $m_{1}, m_{2}, m_{3}$ for the
experimentally allowed classes. Masses are in
eV.}
\end{footnotesize}
\end{center}
\end{small}
\end{table}
\end{landscape}

To this end, in Table 3.3 we have summarized the
allowed ranges of Dirac-like CP violating phase
$\delta$, the Majorana phases $\rho, \sigma$ and
three neutrino masses $m_{1}, m_{2}, m_{3}$ for
the experimentally allowed classes. A few
interesting points are as follows:
\begin{itemize}

\item As is obvious from the table, the matrices
belonging to class A are ruled out for IO,
whereas for NO, although both $A_{1}$ and $A_{2}$
are compatible, however, one is not able to
obtain any useful constraints for either of the
CP violating phases.

\item All the matrices belonging to class B are
viable for both NO and IO. For both the mass
orderings, phase $\delta$ takes values close to
$90^{0}$ and $270^{0}$. The Majorana phases
$\rho$ and $\sigma$ are both quite small, in
particular for IO these take values smaller to
the corresponding values for NO. For NO, these
conclusions have already been shown graphically
in Figs. \ref{fig2} and \ref{fig3}(a), (b). The
corresponding graphs pertaining to IO are similar
and hence not shown here.

\item Again the matrix belonging to class C is viable
for both NO and IO. Unlike the matrices of class
B, now one finds significantly different
constraints on CP violating phase $\delta$ for NO
and IO. In particular, for NO, $\delta$ takes
values from $0^{0}-62.4^{0}$, $114^{0}-244^{0}$
and $294^{0}-360^{0}$. However for IO, the lower
limit of $\delta$ becomes considerable higher, in
particular $\delta$ takes values from
$40.5^{0}-86.8^{0}$, $93.2^{0}-266.5^{0}$ and
$273.2^{0}-325^{0}$. Further, regarding the
Majorana phases, for NO, the phases $\sigma$ as
well as $\rho$ lie between $-90.0^{0}$ to
$-27.3^{0}$ and $27.3^{0}-90.0^{0}$. However, for
IO, although phase $\sigma$ remains
unconstrained, the phase $\rho$ gets considerably
constrained to $-18.4^{0}-18.2^{0}$.

For NO, these conclusions have already been shown
graphically in Figs. \ref{figC} and
\ref{figCphases}(a), (b). For IO, mixing angle
$\theta_{13}$ versus phase $\delta$ graph is
shown in Fig. \ref{figCIO}. From the graph, one
finds that corresponding to the 3$\sigma$ CL
experimental range of angle $\theta_{13}$ i.e.
$7.9^\circ$ to $9.3^{\circ}$, one obtains a
slight constraint on phase $\delta$, in
particular it largely lies between
$40^{0}-325^{0}$.

\begin{figure}[ht]
\begin{center}
\includegraphics[width=0.40\columnwidth]{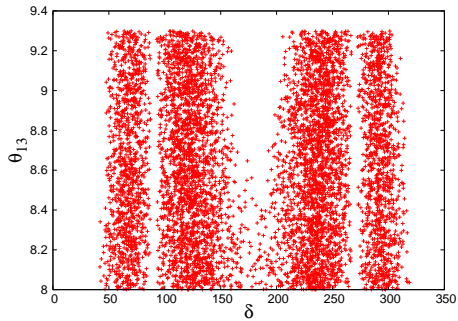}
\caption{Pattern $C$: Mixing angle $\theta_{13}$
versus phase $\delta$ for IO. All the parameters
are in degrees.} \label{figCIO}
\end{center}
\end{figure}

Again for IO, in Fig. \ref{figCphasesIO}(a) and
(b), we have shown the correlation plots of
Dirac-like CP violating phase $\delta$ and
Majorana phases phases $\rho$ and $\sigma$.
Interestingly, we find that although for phase
$\sigma$ we do not obtain any useful constraint,
i.e., it takes values from $-90^{0}-90^{0}$,
however, phase $\rho$ gets narrowed down
considerably, i.e., it lies between
$-18^{0}-18^{0}$. Also, one finds that phase
$\rho$ shows considerable dependence on
Dirac-like CP violating phase $\delta$.

\begin{figure}[ht]
\begin{center}
\subfigure[]{\includegraphics[width=0.40\columnwidth]{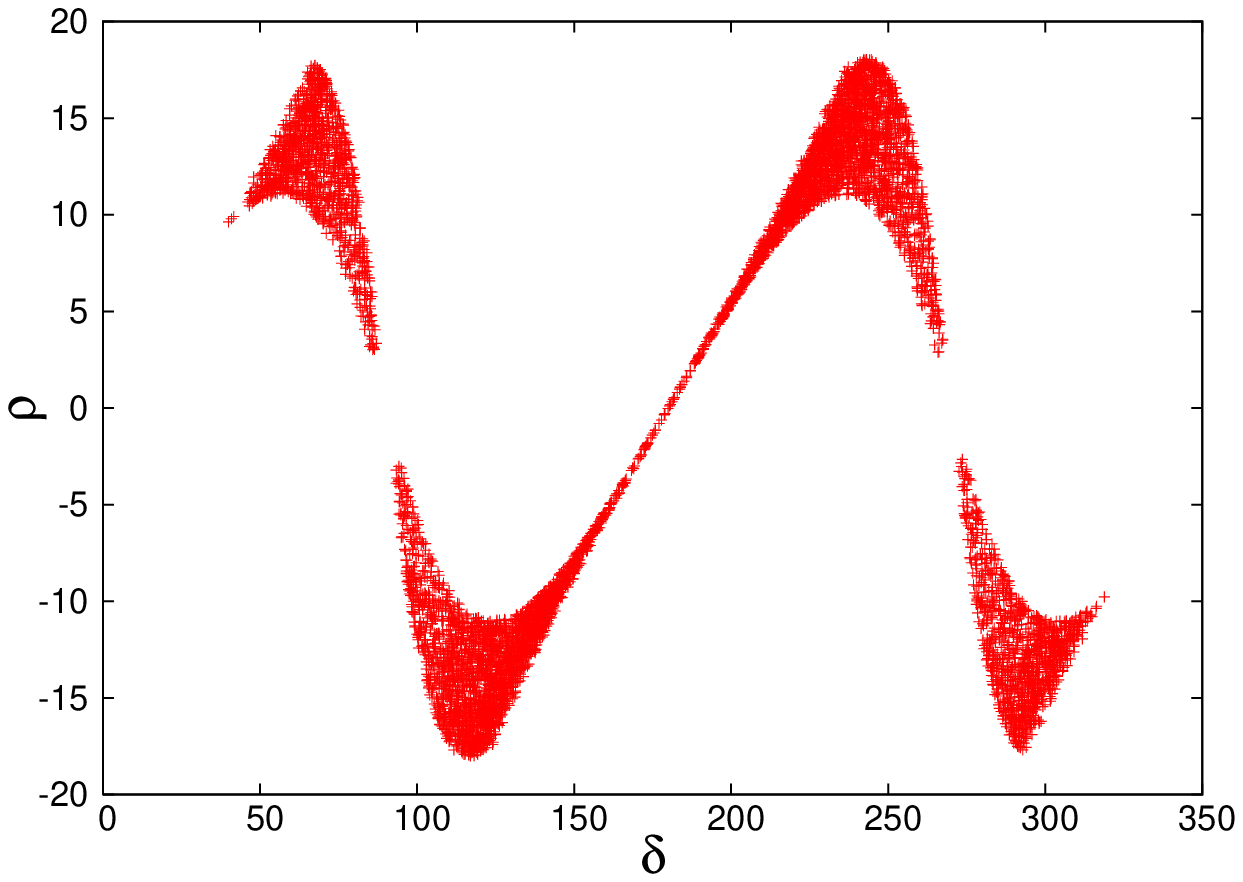}}
\ \ \
\subfigure[]{\includegraphics[width=0.40\columnwidth]{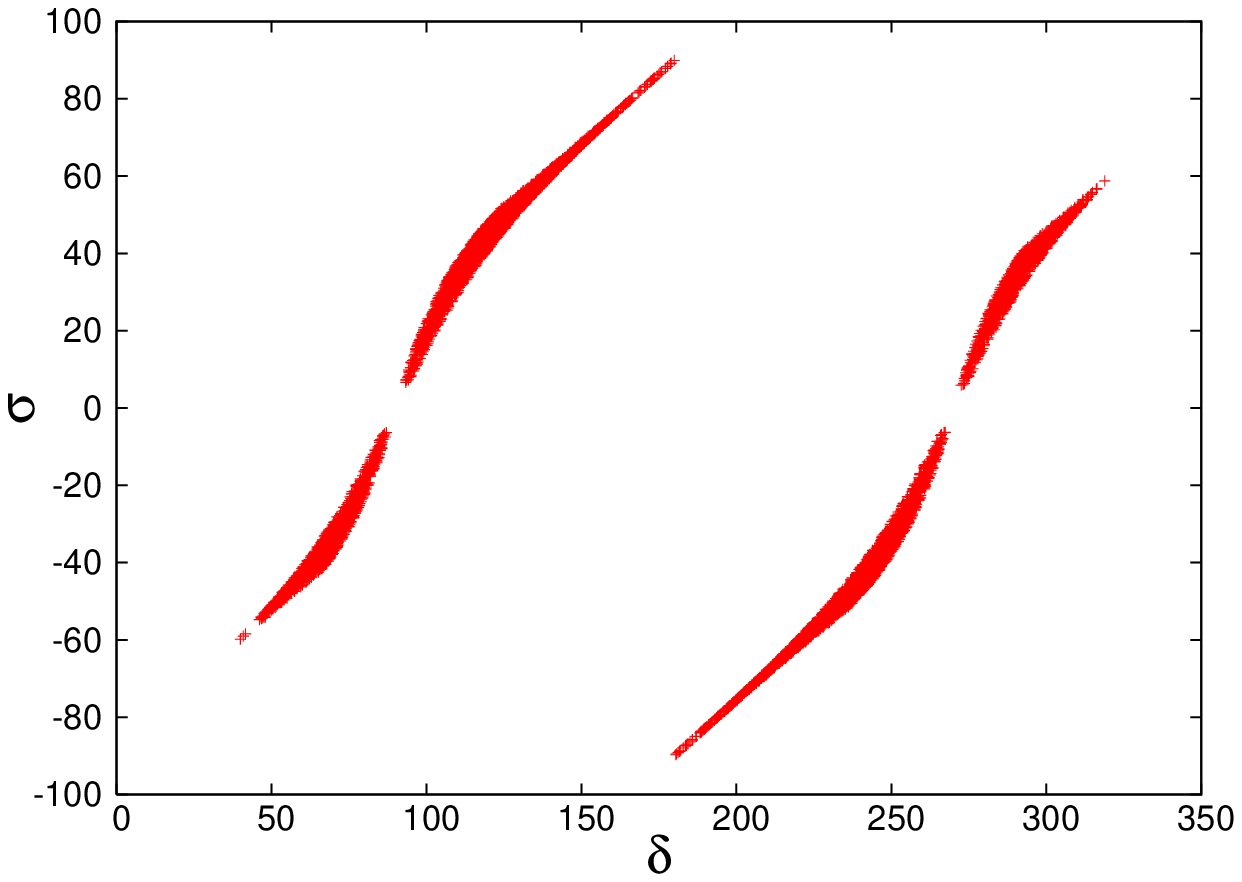}}
\caption{\label{figCphasesIO} Pattern $C$:
Correlation plots of phases $\rho$ and $\sigma$
versus $\delta$ for IO. All the parameters are in
degrees.}
\end{center}
\end{figure}

\end{itemize}

\section{Summary and conclusions}\label{conc}
To summarize, in the light of recent refinements
of mixing angle $\theta_{13}$, we have
re-investigated all the possible texture
structures of three zero and two zero neutrino
mass matrices for both Majorana and Dirac
neutrinos in the flavor basis. In particular, we
have examined implications of angle $\theta_{13}$
on the parameter space of three CP violating
phases $\rho, \sigma$, $\delta$. For the case of
texture three zero mass matrices, in confirmation
with the earlier results, we find that all
possible structures remain ruled out even with
the recent data. For a particular texture
structure, this has been shown using correlation
plots between angles $\theta_{12}$ and
$\theta_{23}$ for the constraints obtained from
real and imaginary parts of ratios of complex
neutrino mass eigenvalues. These graphs show that
$\theta_{12}$ obtained here has no overlap with
its experimental range, therefore, ruling out the
present texture specific case.

For the case of texture two zero mass matrices,
out of the 15 possible structures, the matrices
belonging to classes D, E and F remain ruled out
even with the latest data. For the classes A, B
and C, analysis has been carried out for both NO
and IO. The analysis shows that the matrices of
class A are ruled out for IO, whereas, for NO,
one does not obtain any useful constraints on the
CP violating phases. For the class B matrices,
all these are viable for both NO and IO. For both
the mass orderings, phase $\delta$ takes values
close to $90^{0}$ and $270^{0}$, however, the
Majorana phases $\rho$ and $\sigma$ both acquire
quite small values. The matrix belonging to class
C is again viable for both NO and IO. For NO,
phase $\delta$ has lower limit $0^{0}$, whereas,
for IO the lower limit is $40.5^{0}$. The
Majorana phases also acquire different
constraints for different mass orderings.

\vskip 0.5cm {\bf Acknowledgements} \\ M.S.
acknowledges the Chairperson, Department of
Physics, P.U., for providing facilities to work.
G.A. would like to acknowledge DST, Government of
India (Grant No: SR/FTP/PS-017/2012) for
financial support.  M.G. would like to
acknowledge CSIR, Govt. of India, (Grant
No:03:(1313)14/EMR-II) for financial support.


\end{document}